\documentstyle[sprocl]{article}

\font\eightrm=cmr8

\input{epsf}

\def\be{\begin{equation}}
\def\ee{\end{equation}}
\def\bea{\begin{eqnarray}}
\def\eea{\end{eqnarray}}

\begin{document}

\title{The Dirac Operator Spectrum and Effective Field Theory }

\author{P. H. DAMGAARD}

\address{The Niels Bohr Institute, Blegdamsvej 17, DK-2100 Copenhagen, 
Denmark} 




\maketitle\abstracts{When chiral symmetry is spontaneously broken,
the low-energy part of the Dirac operator spectrum can be computed
analytically in the chiral limit. The tool is effective field theory
or, equivalently in this case, Random Matrix Theory.}

\section{Introduction}

Contrary to what one might naively have expected, it is possible to
predict the analytical behavior of the Dirac operator spectrum in
QCD and QCD-like theories for eigenvalues very close to
zero.\footnote{Plenary talk at ``Confinement IV'', Vienna, July, 2000. To
appear in those proceedings} This is the low-energy regime of the theory, 
and ordinary
QCD perturbation theory is completely irrelevant here. In fact,
the analytical predictions turn out to be not only non-perturbative
in the QCD coupling constant, but {\em exact}. These developments 
began eight years ago with a paper by Leutwyler
and Smilga \cite{LS}, where for the first time it was shown that
if QCD undergoes spontaneous chiral symmetry breaking, there are
constraints on the Dirac operator eigenvalues near the origin.
So the idea is not new, but there have been many new
developments in just the past few months. Here I 
summarize some of the recent work I have been involved in myself. For
a comprehensive review of the subject up to last year, see ref.
\cite{VW}.

What originally pushed this subject forward was the suggestion
by Shuryak and Verbaarschot \cite{SV} that the lowest part of the Dirac
operator spectrum could be computed analytically by means of Random Matrix
Theory. Much work has gone into understanding this remarkable connection
between chiral symmetry breaking in gauge theories and Random Matrix
Theory.
In particular, it has been shown that the chiral (flavor) symmetry
breaking in gauge theories coupled to fermions can be classified precisely
according to the main categories of Random Matrix Theory ensembles 
\cite{V,VZ,M}. In the limit relevant for applications to the Dirac operator
spectrum these Random Matrix Theory ensembles actually provide {\em
universality classes} \cite{ADMN}: results do not depend in detail on the
particular distributions of the random matrices.

Random Matrix Theory should not be required to derive results for the Dirac
operator spectrum; all should follow from field theory alone. It is one
of the more recent discoveries that this is indeed the case. There are
now three independent means of deriving analytical results for
the Dirac operator spectrum:

\begin {itemize}
\item Random Matrix Theory
\item The supersymmetric method
\item The replica method
\end{itemize}

Both the so-called supersymmetric \cite{OTV} and replica \cite{DS,DV}
methods stand out, as they rely only on the field theory formulation itself.
Because results are claimed to be {\em exact}, all three methods should
yield identical results. This is indeed the case. One pleasing 
consequence is that one can now use any of the three different formulations
according to what is most convenient. Generically the Random Matrix Theory
formulation is the easiest for analytical calculations, but there are
cases where it is actually simpler to use the effective field theory
partition function \cite{AD} (see also refs. \cite{NN,AK}). 
Even in those cases, the explicit derivations
have usually gone through Random Matrix Theory at intermediate steps.

In this short review I will cover three different topics on which there
has been progress in just the last few months. These are: ({\bf i}) the 
analytical computation of individual eigenvalue distributions \cite{DN} 
(and subsequent lattice measurements \cite{DHNR}), ({\bf ii}) the replica 
method as applied to the finite-volume effective partition function of QCD
\cite{DS,DV}, and ({\bf iii}) lattice
Monte Carlo measurements of the low-energy end of the Dirac operator spectrum
near the chiral phase transition \cite{FF,DHNR1}.

\section{Distributions of individual Dirac operator eigenvalues}

Analytical expressions for the distribution of the first non-zero
Dirac operator eigenvalue have been known for some time in the case
of all three chiral ensembles. Recently a much more general analytcial
expression was found for the probability distribution of the
$k$th smallest Dirac operator eigenvalue (again counted above the
exact zero modes) \cite{DN}. The formula is completely general, holds
for any number of (possibly massive) fermions, and for any of the three
major chiral universality classes, as labelled by their Dyson index
$\beta$. The only technical restriction is that for the universality class of
$\beta=1$ (corresponding to gauge group SU(2) and fermions in the
fundamental representation), the formula only works for a sector of
{\em odd} topological charge $\nu$. While the analytical expression 
has been derived
in the framework of Random Matrix Theory, the final formula turns out
to depend only on the effective field theory partition functions. It
generalizes a related formula for just the lowest eigenvalue distribution
in the case of the $\beta=2$ universality class, which also expressed
the distribution in terms of finite volume partition functions \cite{NDW}.

Before presenting the resulting closed analytical expression for the
distribution of the $k$th smallest Dirac operator eigenvalue, let us first 
recall some basic facts about the connection to Random Matrix Theory.
Because chiral symmetry is presumed spontaneously broken, the large-volume
euclidean partition function is, in the chiral limit, entirely dominated
by the pseudo-Goldstone bosons. The dominance is very strong: contributions
from all other physical excitations are exponentially suppressed
in the masses, with a coefficient in the exponent that grows linearly
with the size of the box. As the size of the box is sent to infinity,
the Goldstone bosons dominate the euclidean partition function entirely.
It is in this sense that the results for the Dirac operator eigenvalues
are exact: they can be made to reach any desired accuracy by taking
the volume large enough, and the quark masses small enough. By imposing 
on the four-volume that $V \gg m_{\pi}^{-4}$ only the zero
mode piece of the Goldstone field $\pi(x)$ survives, and the effective
partition function, in a sector of topological charge $\nu$ becomes
\cite{LS}
\be
{\cal Z}_{\nu}^{(N_{f})}(\{\mu_i\}) ~=~ \int_{U(N_f)}\! dU~ (\det U)^{\nu}
\exp\left[\frac{1}{2}V\Sigma {\mbox{\rm Tr}}({\cal M}U^{\dagger}
+ U{\cal M}^{\dagger})\right] ~.\label{ZLS}
\ee
This partition function, a zero-dimensional group integral has, surprisingly, 
a Random Matrix Theory representation \cite{SV}. To get it, 
one starts with \cite{SV}
\be 
\tilde{\cal Z}_{\nu}^{(N_{f})}(\{m_i\}) 
~=~ \int\! dW \prod_{f=1}^{N_{f}}{\det}\left(iM + m_f\right)~
\exp\left[-\frac{N}{2} \mbox{tr}\, V(M^2)\right] ~,\label{ZRMTchUE}
\ee
where
\be
M ~=~ \left( \begin{array}{cc}
              0 & W^{\dagger} \\
              W & 0
              \end{array}
      \right) ~.
\ee
The complex matrix $W$ in eq. 
(\ref{ZRMTchUE}) is of size $(N+\nu)\times N$, and the potential is
essentially not constrained beyond yielding a non-vanishing spectral density
at the origin, $\rho(0) \neq 0$ \cite{ADMN}. The limit $N\to \infty$ is 
taken in the Random Matrix Theory partition function
in such a way that the combinations $\mu_i \equiv m_i\pi\rho(0)2N$,
are kept fixed. In this limit the partition functions
${\cal Z}_{\nu}^{(N_f)}$ and $\tilde{\cal Z}_{\nu}^{(N_{f})}$ become
equal up to a $\mu_i$-independent constant \cite{SV}.

For practical computations an eigenvalue representation of the above
matrix integral is convenient. Taking a general Dyson index $\beta
=1, 2$ or 4, it
can, up to an overall and irrelevant normalization factor, be written as
\be
Z^{(\beta)}_\nu =
\prod_{i=1}^{N_f} m_i^\nu
\int_0^\infty
\prod_{i=1}^N 
\Bigl(dx_i\, x_i^{\beta(\nu+1)/2-1}
e^{-\beta x_i}
\prod_{j=1}^{N_f} (x_i+m_j^2) \Bigr)
\prod_{i > j}^N |x_i-x_j|^\beta ~, \label{Zeigenv}
\ee
where $x_i$ are the eigenvalues of $M^2$.
Here the potential has been chosen to be just $V(x) = x$. This is precisely
permitted because of universality of the final results \cite{ADMN}. Due to
symmetry under $\nu \to -\nu$ it is 
convenient to restrict oneself to $\nu \geq 0$.

The unnormalized joint probability distribution
for all $N$ eigenvalues is
\be
\rho^{(\beta)}_N (x_1,\ldots,x_N;\{m_i^2\})=
\prod_{i=1}^N \Bigl(x_i^{\beta(\nu+1)/2-1}
e^{-\beta x_i} \prod_{j=1}^{N_{f}} (x_i+m^2_j) \Bigr)
\prod_{i > j}^N |x_i-x_j|^\beta ~.
\ee
Similarly, the unnormalized joint probability distribution of
the $k$ smallest eigenvalues
$\{0\leq x_1\leq \cdots \leq x_{k-1}\leq x_k\}$
is given by
\bea
&&{\Omega}^{(\beta)}_{N,k}(x_1,\ldots,x_k;\{m_i^2\})
=
\int_{x_{k}}^\infty dx_{k+1} \cdots dx_{N}\,
\rho^{(\beta)}_N (x_1,\ldots,x_N;\{m^2\})
\nonumber\\
&&=
\prod_{i=1}^k \Bigl(x_i^{\beta(\nu+1)/2-1} e^{-\beta x_i}
\prod_{j=1}^{N_{f}} (x_i+m^2_j) \Bigr)
\prod_{i>j}^k |x_i-x_j|^\beta
\nonumber\\
&&\times
\int_{x_{k}}^\infty dx_{k+1} \cdots dx_{N}\,
\prod_{i=k+1}^N \Bigl(x_i^{\beta(\nu+1)/2-1} e^{-\beta x_i}
\prod_{j=1}^{N_{f}} (x_i+m^2_j)
\prod_{j=1}^{k} (x_i-x_j)^\beta \Bigr)\cr
&&\times \prod_{i >j\geq k+1}^N |x_i-x_j|^\beta ~,
\label{Omega}
\eea

Shifting $x_i \to x_i + x_k$ in the integrand gives
\bea
&&{\Omega}^{(\beta)}_{N,k}(x_1,\ldots,x_k;\{m_i^2\})
~=~ e^{-(N-k)\beta x_k}
\prod_{i=1}^k \Bigl(x_i^{\beta(\nu+1)/2-1} e^{-\beta x_i}
\prod_{j=1}^{N_{f}} (x_i+m^2_j) \Bigr)\cr
&&\times \prod_{i>j}^k |x_i-x_j|^\beta \int_{0}^\infty 
\prod_{i=k+1}^N \Bigl(dx_{i}\, e^{-\beta x_i}
x_i^{\beta}(x_i + x_k)^{\beta(\nu+1)/2-1}
\prod_{j=1}^{N_{f}} (x_i+m^2_j+x_k) \cr 
&&\times \prod_{j=1}^{k-1}(x_i + x_k -
x_j)^{\beta}
\Bigr)
\prod_{i>j\geq k+1}^N |x_i-x_j|^{\beta}~.
\label{shifted}
\eea
To finally get the probability distributions of the Dirac operator eigenvalues
one takes the microscopic limit $N \to \infty$ with, in this particular 
convention, 
$\zeta_i=\pi \rho(0) \sqrt{x_i} = \sqrt{8N\, x_i}$ and 
$\mu_j=\pi \rho(0) m_j  =\sqrt{8N}m_j$ kept fixed.
In this large-$N$ limit the difference between partition functions
based on $N-k$ and $N$ eigenvalues disappears. One immediately sees
that the new terms in the integrand of (\ref{shifted}) can
be interpreted as arising from new additional fermion species, with
the partition function now being evaluated in a fixed topological sector of
effective charge $\nu = 1 + 2/\beta$. The only restriction is that
for $\beta=1$ the topological charge $\nu$ must be odd (because otherwise
the number of additional fermion species will be fractional).

Taking into
account the definition (\ref{Zeigenv}), this gives:
\bea
&&\omega_k^{(\beta)}(\zeta_1,\ldots,\zeta_k; \{\mu_i\})
~=~ \lim_{N\to \infty} \Bigl(\prod_{i=1}^k \frac{|\zeta_i|}{8N} \Bigr)
{\Omega}_{N,k}^{(\beta)}
(\frac{\zeta_1^2}{8N},\ldots,\frac{\zeta_k^2}{8N};
\{\frac{\mu^2}{8N} \})
\nonumber\\
&&=
C e^{-\beta \zeta_k^2/8}
\zeta_k^{\beta\frac{(\nu+1)}{2}-\nu-1+\frac{2}{\beta}}\prod_{j=1}^{N_{f}}
(\mu_j^2+\zeta_k^2)^{\frac{1}{2} - \frac{1}{\beta}}
\prod_{i=1}^{k-1}\Bigl(\zeta_i^{\beta(\nu+1)-1}
(\zeta_k^2-\zeta_i^2)^{\frac{\beta}{2}-1}\times
\cr &&\!\!\!
\prod_{j=1}^{N_{f}}(\zeta_i^2+m_j^2)
\Bigr)\!\prod_{i>j}^{k-1}(\zeta_i^2-\zeta_j^2)^{\beta}
\prod_{j=1}^{N_{f}}\mu_j^{\nu} 
\frac{{\cal Z}^{(\beta)}_{1+2/\beta}(
\{\sqrt{\mu_i^2+\zeta_k^2}\},
\{\sqrt{\zeta_k^2-\zeta_i^2}\},
\{\zeta_k\})}
{{\cal Z}^{(\beta)}_{\nu}(\{\mu_i\})} 
\label{bboxOmega}
\eea
In the partition function in the numerator each of the $N_f$ fermion
masses have been shifted according to $\mu_i^2 \to \mu_i^2 + \zeta_k^2$.
There are $\beta(k-1)$ new masses $\zeta_k^2 - \zeta_i^2$, $i = 1,
\ldots, k-1$, each of them being $\beta$ times degenerate. Finally
there are also $\beta(\nu+1)/2-1$ new fermions, all of mass $\zeta_k$.
The overall normalization factor $C$ is fixed by the requirement
of probability conservation.

To get the individual eigenvalue distribution of the $k$th eigenvalue,
one simply integrates out the previous $k-1$ smaller eigenvalues, $viz.$,
\be
{p}^{(\beta)}_{k}(\zeta;\{\mu_i\})=
\int_0^\zeta d\zeta_1 \int_{\zeta_{1}}^\zeta d\zeta_2
\cdots \int_{\zeta_{k-2}}^\zeta d\zeta_{k-1}\,
\omega^{(\beta)}_{k}(\zeta_1,\ldots,\zeta_{k-1},\zeta;\{\mu_i\})~ .
\label{p}
\ee

The general formulas (\ref{bboxOmega}) and (\ref{p})  may look rather 
complicated, but they actually simplify considerably in a number of 
interesting situations. For instance, in the physically most interesting
case of QCD (which belongs to the $\beta=2$ universality class \cite{V}),
we get in a sector of topological charge $\nu=0$:
\begin{eqnarray}
&&\omega_k(\zeta_1,\ldots,\zeta_k; \{\mu_i\})
~=~ C e^{-\zeta_k^2/4}
\zeta_k
\prod_{i=1}^{k-1}\Bigl(\zeta_i
\prod_{j=1}^{N_{f}}(\zeta_i^2+\mu_j^2)
\Bigr)\prod_{i>j}^{k-1}(\zeta_i^2-\zeta_j^2)^2\times \cr
&&
\frac{{\cal Z}_{2}\left(
\left\{\sqrt{\mu_i^2+\zeta_k^2}\right\},
\sqrt{\zeta_k^2-\zeta_1^2},\sqrt{\zeta_k^2-\zeta_1^2},
\ldots,
\sqrt{\zeta_k^2-\zeta_{k-1}^2},\sqrt{\zeta_k^2-\zeta_{k-1}^2}\right)}{
{\cal Z}_{0}(\{\mu\})} 
\label{bboxOmega1}
\end{eqnarray}
The finite-volume partition functions involved here are known in closed
analytical form \cite{JSV},
\be
{\cal Z}_{\nu}(\{\mu_i\}) ~=~ \det A(\{\mu\})/\Delta(\mu^2) ~,
\ee
where the determinant is taken over the $N_f\times N_f$ matrix
\be
A(\{\mu\}) ~=~ \mu_i^{j-1}I_{\nu+j-1}(\mu_i) ~,
\ee
and 
\be
\Delta(\mu^2) ~=~ \prod_{i>j}^{N_{f}}(\mu_i^2-\mu_j^2) ~.
\ee
With this convention the normalization factor is $C = 1/2$ for all
values of $k$, $N_f$ and $\nu$.

The analytical formula (\ref{bboxOmega1}) has very recently been tested
by lattice gauge theory simulations \cite{DHNR}, with quite remarkable
agreement even at relatively small volumes. These simulations were
done with staggered fermions, which are almost totally insensitive
to gauge field topology at the couplings we are concerned with here
\cite{DHNR2}. This means that comparisons should be done only with
the $\nu=0$ analytical predictions. Shown in figure 1 is the
result of a large-statistics computation in quenched SU(3) gauge theory.
The individual Dirac operator eigenvalues indeed do have distributions
that fall right on top of the analytical predictions.


\vspace{0.6cm}

\centerline{
{\epsfxsize=9cm\epsfbox{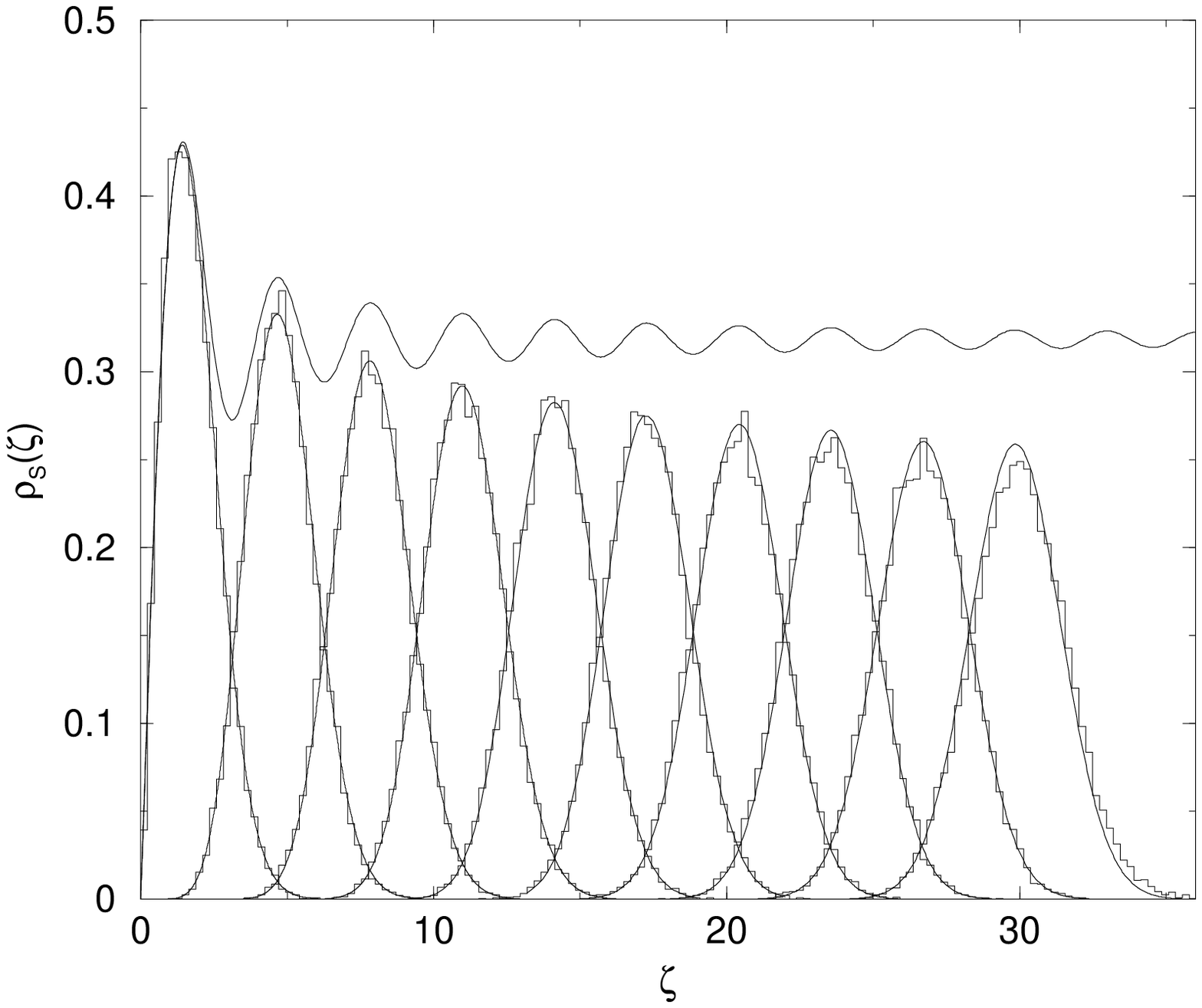}}}
\noindent{\eightrm Fig. 1~~Individual Dirac operator eigenvalue 
distributions from lattice Monte Carlo simulations in quenched
SU(3) gauge theory \cite{DHNR}, and
the analytical predictions of eq. (\ref{bboxOmega1}).}

Simulations with dynamical fermions in SU(3) gauge theory display the
same degree of accuracy (although for obvious reasons the statistics here
is much lower) \cite{DHNR}. Recall that both masses $m_i$ and eigenvalues
are rescaled with the same factor of $\Sigma V$. This means that to get
as good accuracy as possible one needs to probe the theory of nearly
massless quarks. In figure 2 is shown an analogous plot for the theory
with $N_f=1$ (strong coupling) staggered fermions (at a mass value of 
$m=0.003$ and a volume of $6^4$). 

\vspace{0.6cm}

\centerline{
{\epsfxsize=9cm\epsfbox{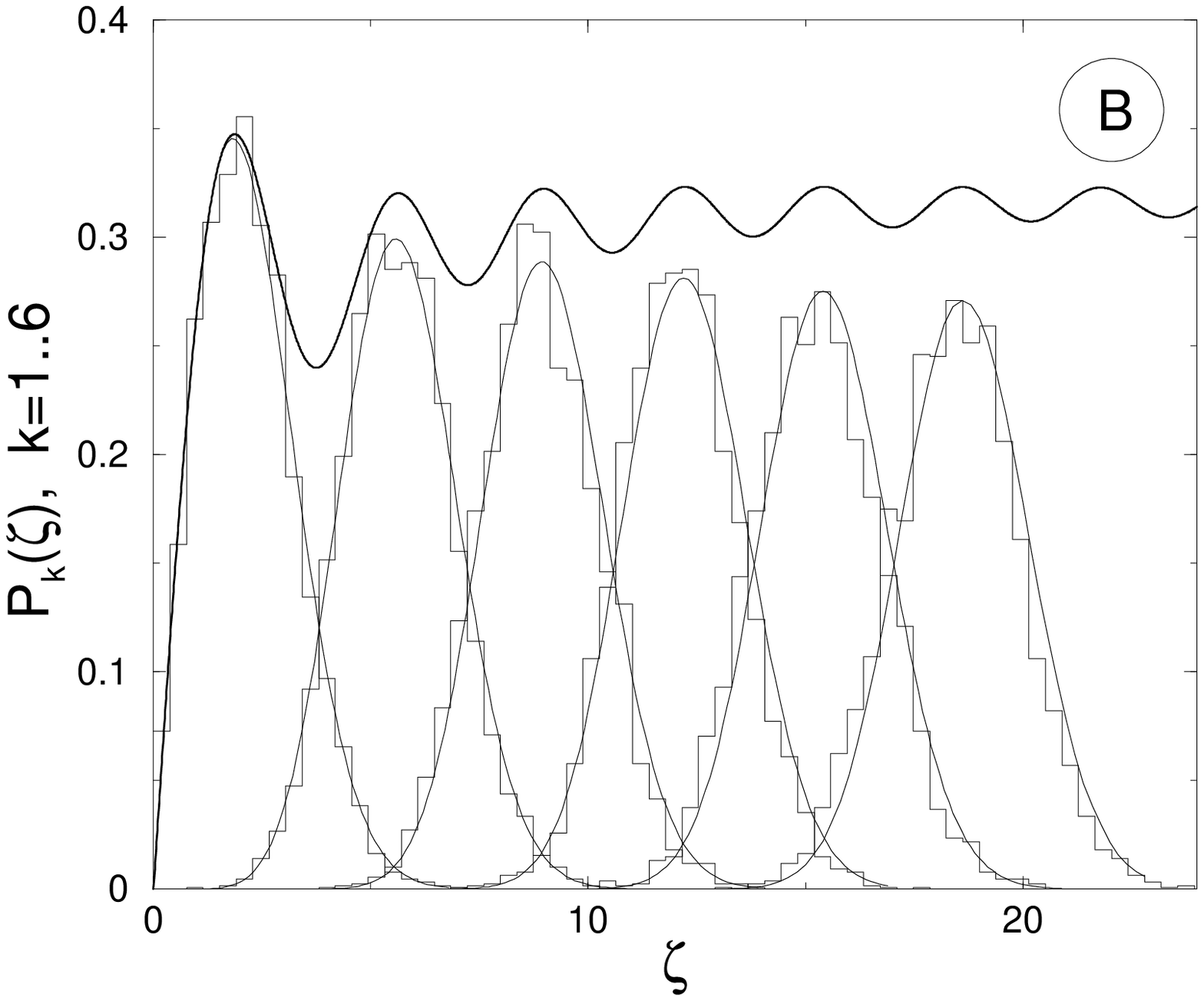}}}
\noindent{\eightrm Fig. 2~~Individual Dirac operator eigenvalue 
distributions from lattice Monte Carlo simulations of SU(3) gauge theory
with $N_f=1$ \cite{DHNR}, compared with
the analytical predictions of eq. (\ref{bboxOmega1}).}

One sees the same kind of detailed 
agreement as in the quenched theory. This establishes quite clearly
that not only can the microscopic spectral density of the Dirac operator
be computed analytically in the relevant scaling window, also individual
Dirac operator eigenvalues are falling right on the analytical
predictions.

\section{The replica method}

In lattice gauge theory simulations one often starts out with quenched
calculations: the fermion determinant is entirely ignored in the averages.
This is just a simple (sometimes accurate, sometimes not) approximation,
introduced only in order to save computer time. 
A more sophisticated approach is ``partial quenching''. Here one does
simulate the full theory with dynamical fermions, but in addition one extracts
{\em more} information from the same configurations. This is done by
computing quenched averages in the field configurations that already
included the effects of dynamical fermions. At first sight it may sound
absurd to do this, but in fact there is much genuine physics to be
extracted from such partially quenched simulations. In addition, it turns
out that at the analytical level partial quenching is just what is needed
in order to derive properties of the Dirac operator spectrum. In this
case partial quenching is not at all considered as an approximation to the
full theory; rather, it is used to obtain genuine physical information
about the real theory, the one with propagating dynamical quarks.

To make a partially quenched average, we must measure some correlation
function of new fictituous ``valence'' quarks that do not affect 
the configurations. The perhaps most simple quantity is the partially quenched
chiral condensate. We can get this condensate by adding $N_v$ valence
quarks, differentiate with respect to their mass, and subsequently taking
the limit $N_v\to 0$:
\bea
\frac{\Sigma_{\nu}(\mu_v,\{\mu_i\})}{\Sigma} &\equiv&
\lim_{N_{v}\to0}\frac{1}{N_{v}}\frac{\partial}{\partial\mu_{v}}
\ln{\cal Z}^{(N_{f}+N_{v})}_{\nu}(\mu_v,\{\mu_i\}) \cr
&=& \left.\frac{1}{{\cal Z}^{(N_{f}+N_{v})}_{\nu}(\mu_v,\{\mu_i\})}
\frac{\partial}{\partial N_{v}}
\frac{\partial}{\partial\mu}{\cal Z}^{(N_{f}+N_{v})}_{\nu}
(\mu_v,\{\mu_i\})\right|_{N_{v}=0} 
\label{Sigmadef}
\eea
In the last line we have formally expanded the partition function as
a Taylor series in $N_v$. It is not obvious that  
precise meaning can be given to such a notion, since the function
involved {\em a priori} is known only at integer values of $N_v$. 

This particular way of deriving the partially quenched chiral condensate
is known as the replica method. Its applicability in certain condensed
matter contexts has been questioned \cite{VZirn}, but last year there was
considerable progress towards understanding how to apply this method
to derive spectral properties \cite{Mezard}. In the present context of
QCD, the partially quenched chiral condensate in the same finite-volume
scaling regime as discussed above was computed in refs. \cite{DS,DV}.  
The replica method trivially works in ordinary QCD perturbation theory
(it simply kills all closed fermion loops, as required). It is a more
non-trivial fact that the replica method also is suited for deriving
quantities that are non-perturbative in the QCD coupling constant.
For instance, in the low-energy framework of effective chiral Lagrangians
the replica method \cite{DS1} works quite analogous to the previously
known supersymmetric method \cite{BG}.

Although in principle the partially quenched chiral condensate is an
unphysical quantity, it turns out that this is not {\em quite} so. There
is one tiny bit of this quantity that contains important physics: this
the discontinuity across the imaginary mass axis \cite{OTV}, which 
gives the spectral density. In the microscopic scaling region:
\bea
\rho_S^{(\nu)}(\zeta;\{\mu_i\}) &=& \frac{1}{2\pi} {\mbox{\rm Disc}}
\left.\Sigma_{\nu}(\mu_v,\{\mu_i\})\right|_{\mu_v=i\zeta} \cr
&=& \frac{1}{2\pi}[\Sigma_{\nu}(i\zeta+\epsilon;\{\mu_i\}) - 
\Sigma(i\zeta-\epsilon;\{\mu_i\})] ~.
\label{rhodisc}
\eea
In ref. \cite{DS} two expansions of the partially quenched chiral condensate
were considered: small-mass and large-mass expansions. It turned out
that neither were suitable for deriving the spectral density of the
Dirac operator. The small-mass expansion suffered from so-called
de Wit--`t Hooft poles \cite{DeW}, while the large-mass expansion of
ref. \cite{DS}, based as it were on the leading saddle point, only
gave the exponentially leading asymptotic series near the real axis;
it could not be trusted near the imaginary axis, as required to get
the spectral density.
Quite recently, Dalmazi and Verbaarschot \cite{DV} have shown how to
repair this latter deficiency of the large-mass expansion, by including
other saddles. Indeed, their asymptotic expansion for large masses is
valid also near the imaginary axis. Taking the discontinuity there 
according to eq. (\ref{rhodisc}) precisely reproduces the asymptotic
expansion of the mircoscopic spectral density of the Dirac operator
as computed based on the Random Matrix Theory formulation. This is
a highly non-trivial fact, since in detail the computations are entirely
different from those of either the Random Matrix Theory context or
the supersymmetric formulation. Conceptually, the replica method has
the advantage that the pattern of chiral symmetry breaking safely
can be assumed to be the conventional one: SU($N_f+N_v)_L\times$ 
SU($N_f+N_v)_R \to$ SU($N_f+N_v$), whereas in the supersymmstric
formulation the symmetry breaking pattern is simply assumed to mimic as
closely as possible the known bosonic one. The corresponding difficulty
in the replica method is obviously how this pattern of chiral symmetry
breaking can be given meaning for non-integer $N_v$ (or rather, infinitesimal
$N_v$, which is all that is required). The fact that the replica method
happens to agree with the supersymmetric one can, since they are 
so entirely different, be seen as an independent confirmation of both.

\section{Smallest Dirac operator eigenvalues near $T_c$}

Because of the Banks-Casher relation between the infinite-volume chiral
condensate $\Sigma$ and the spectral density of the Dirac operator at
the origin, $\Sigma = \pi \rho(0)$, it is obviously of interest to 
trace the depletion of Dirac eigenvalues at finite temperature. A lattice
study of this question was first performed in ref. \cite{FF}, and this
spring similar issues were addressed with higher statistics \cite{DHNR1}.

The obvious question to ask is: can the Random Matrix Theory formulation
of the effective Lagrangian (\ref{ZLS}) be used to predict the Dirac
operator spectrum near $T_c$? The answer is clearly in the negative,
since the effective finite-volume chiral Lagrangian (\ref{ZLS}) simply
is incorrect at finite temperature. The best one can do is to take into
account the leading effect of replacing the symmetric euclidean four-volume
$V = L^4$ by an asymmetric one, $V=L^3/T$, where $T$ still is on the order
of $1/L$. This means that this effective chiral Lagrangian is suitable
only for probing infinitesimally small temperatures, and the same goes
for the associated Random Matrix Theory formulation.

This does not mean that it is uninteresting to study the behavior of
the smallest Dirac operator eigenvalues as the temperature is increased
towards the critical temperature $T_c$ of the chiral phase transition.
For instance, one observation of ref. \cite{DHNR1} was that tracing
the evolution in the magnitude of just the single smallest Dirac operator
eigenvalue is a remarkably simple way to get an accurate detrmination
of the phase transition point. Shown in figure 3 is an example of this,
taken from ref. \cite{DHNR1}.

\vspace{0.4cm} 

\centerline{
{\epsfxsize=9cm\epsfbox{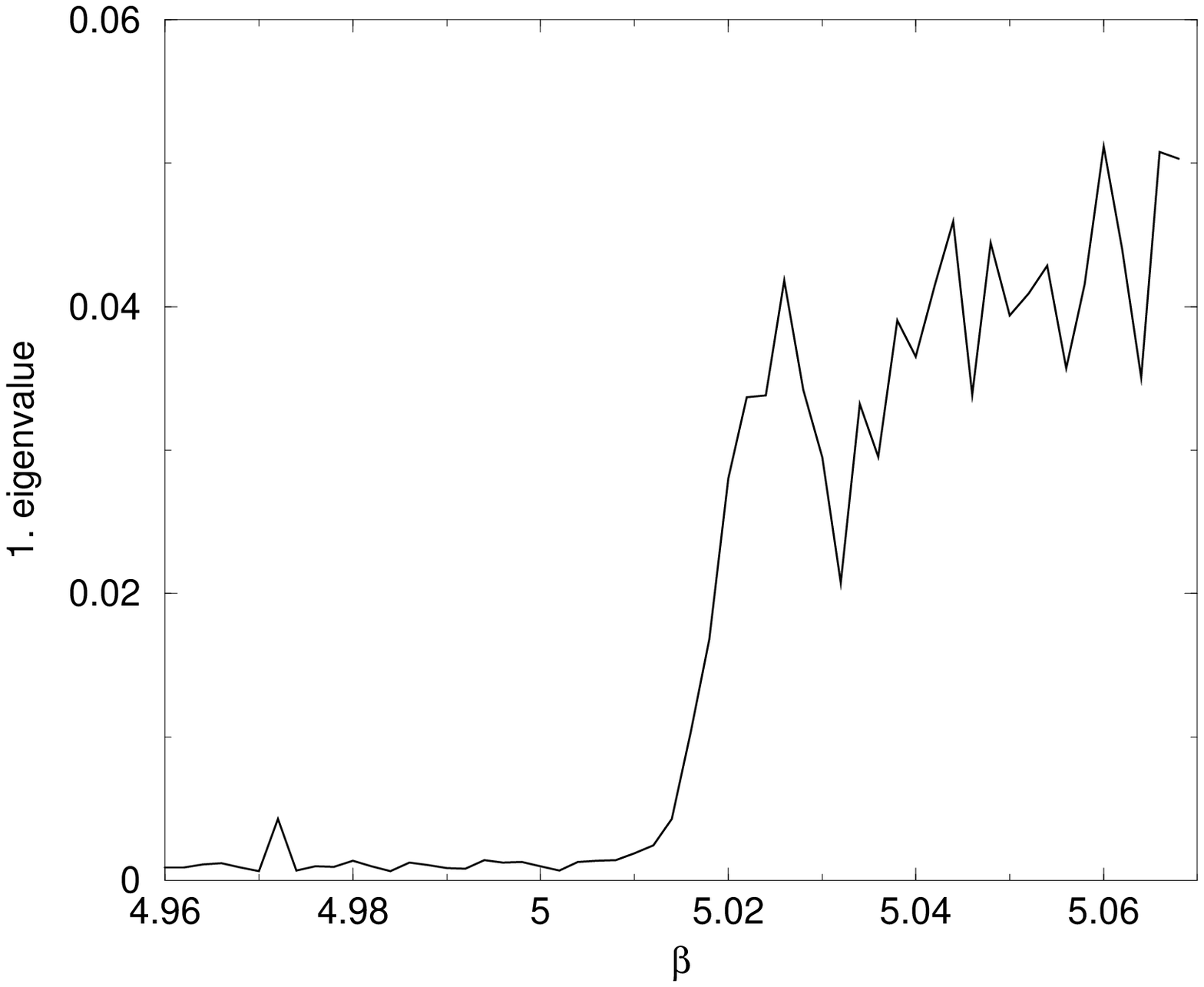}}}
\noindent{\eightrm Fig. 3~~ Just one single Dirac operator eigenvalue,
traced through a series of ensembles with almost no averaging,
suffices for getting an accurate determination of the coupling
at which the chiral phase transition occurs. From ref. \cite{DHNR1}.}

A more ambituous goal would be to try to compute the corresponding
microscopic spectral density at, say, precisely the critical temperature
$T_c$. (For this to make sense, one should need a number of fermuions  $N_f$
for which the transition is continuous). The procedure is as clear as in
the case of zero temperature, but a systematic way of obtaining the
effective theory is lacking. Presumable the closest one can get is an
effective Lagrangian framework like that 
of Pisarski and Wilczek \cite{PW}. This concerns,
however, an effective theory of which not even the leading behavior
as a function of mass is known analytically. Progress on this front is more
likely to come solely from the numerical side.

\vspace{0.3cm}

\section*{Acknowledgments}
Thanks go to U. Heller, R. Niclasen, S. Nishigaki,
K. Rummukainen and K. Splittorff; 
the work described here was all done in collaboration
with them. In addition, I thank G. Akemann, D. Dalmazi and
J.J.M. Verbaarschot for discussions. This work was partially
supported by EU TMR grant no. ERBFMRXCT97-0122.

\end{document}